# Dynamic Systems Model for Ionic Mem-Resistors based on Harmonic Oscillation

Blaise Mouttet

*Abstract*— Memristive system models have previously been proposed to describe ionic memory resistors. However, these models neglect the mass of ions and repulsive forces between ions and are not well formulated in terms of semiconductor and ionic physics. This article proposes an alternative dynamic systems model in which the system state is derived from a second order differential equation in the form of a driven damped harmonic oscillator. Application is made to Schottky and tunneling barriers.

*Keywords- mem-resistor, non-linear dynamic systems*

## I. INTRODUCTION

In 1971 the "memristor" had been proposed by Leon Chua as a "missing" fundamental circuit element exhibiting charge-dependent resistance [1] and was recently claimed to be "found" by researchers of HPLabs in the form of $TiO_{2-x}$ thin films [2]. In actuality there were earlier physical examples of charge-dependent memory resistors [3] and it had been demonstrated since the 1960's that $TiO_2$ thin films exhibit a memory resistance effect [4]. Commercial interest in memory resistors has gradually been increasing since the 1990's and has been under development in numerous forms by AMD, Axon Technologies, Micron Technologies, Samsung, Sharp, Unity Semiconductor, and other companies with applications directed toward non-volatile memory [5]. The recent efforts of HPLabs to market Chua's "memristor" as a newly discovered "$4^{th}$ fundamental circuit element" are arguably an attempt to take credit for the work of these other companies rather than representative of any new discovery [6].

The researchers at HPLabs have made some attempts to model their $TiO_2$ memory resistor using ionic drift/diffusion of oxygen vacancies [7] formulated in terms of memristive systems [8]. However, the memristor ionic drift model is at minimum incomplete (if not entirely incorrect) since:

a) it neglects hysteretic capacitive effects that would occur as ions move in a thin semiconductor film,

b) it neglects the mass and inertial effect of the mobile ions,

c) it neglects repulsive forces between ions, and

d) it neglects dynamic effects of the built-in voltage.

In order to incorporate the above effects the memristor model would need to be modified (or completely replaced) with a different dynamic systems framework. There have been some suggestions of broader dynamic systems models such as mem-admittance systems (Fig. 1b) which incorporate memory capacitive effects found in certain perovskite and nanocrystal films [9]. In addition mem-transistor systems (Fig. 1c) have been suggested to model 3-terminal memory devices such as the Widrow-Hoff memistor or synaptic floating gate memory cells [10]. Yet another conceivable type of mem-electronic system is mem-active systems (Fig.1d) in which an active source of voltage or current is included in a resistance memory cell and exhibits independent memory effects. One example of this type of mem-electronic system may be found in a thin film having a charge storing medium providing non-volatile voltage storage [11]. Another example of mem-active systems may be rechargeable batteries which are resistive when acting as an electrolytic cell but serve as a power source when acting as a galvanic cell.

The present article presents a new dynamic model which incorporates aspects of these different mem-electronic systems to properly address the physics of nano-ionic thin films.

Memristive Systems (Fig.1a):

$$y(t) = g(\mathbf{x}, u, t)u(t)$$
$$\frac{d\mathbf{x}(t)}{dt} = f(\mathbf{x}, u, t) \quad (1)$$

Mem-Admittance System (Fig.1b):

$$y(t) = g(\mathbf{x}, u, t)u(t) + \frac{d}{dt}(c(\mathbf{x}, u, t)u(t))$$
$$\frac{d\mathbf{x}(t)}{dt} = f(\mathbf{x}, u, t) \quad (2)$$

Mem-Transistor System (Fig. 1c):

$$y_1(t) = g(\mathbf{x}, u_1, u_2)$$
$$y_2(t) = h(\mathbf{x}, u_1, u_2)$$
$$\frac{d\mathbf{x}(t)}{dt} = f(\mathbf{x}, u_1, u_2) \quad (3)$$

Mem-Active System (Fig. 1d):

$$y(t) = g(\mathbf{x}, u, t)(u(t) - u_i(\mathbf{x}, u, t))$$
$$\frac{d\mathbf{x}(t)}{dt} = f(\mathbf{x}, u, t) \quad (4)$$

## II. IONIC DYNAMICS IN THIN FILMS

The assumption by [8] that the dynamics of ionic motion can be treated solely by drift-diffusion equations is flawed. It neglects the inertia of ions in a solid and the repulsive forces between ions. A better model for ion dynamics in a solid is provided by Bisquert et al. [12]. This model expresses ionic dynamics in terms of Newton's 2$^{nd}$ law of motion (5) relating the acceleration $d^2x/dt^2$ of an ion having effective mass $m_{ion}$ to the sum of the forces $F_i$ acting on it.

$$m_{ion}\frac{d^2x}{dt^2} = \sum_i F_i \quad (5)$$

When an external electric field is applied to an ionic thin film sandwiched between two electrodes there are three principle forces which act on the ion. The first force ($F_c$) is due to collisions of the ion as it moves through the thin film. The product of this force and the average time between collisions $\tau_c$ can be equated to the change in the ion momentum.

$$F_c \tau_c = -m_{ion}\frac{dx}{dt} \quad (6)$$

The second force ($F_r$) is due to the internal electric field ($E_r$) produced by other ions in the thin film. Assuming an equilibrium position $x_0$ where the forces of the other ions all balance and an ion donor density of $N_d$ this force may be expressed using Gauss's Law as (7).

$$F_r = zeE_r = -\frac{(ze)^2 N_d}{\varepsilon_r \varepsilon_0}(x - x_0) \quad (7)$$

where $\varepsilon_0$ is the vacuum permittivity, $\varepsilon_r$ is the relative permittivity, $e$ is the unit charge, and $z$ is the valence of the ion.

The third force $F_a$ is that produced by an applied voltage bias. According to the ion hopping model reviewed in [13] the ionic current (i.e. $dx/dt$) dependence on the external field is given by a sinh function which is approximated by an exponential at high fields and a linear function at low fields. It may thus be inferred from (6) that the ionic driving force would take the form

$$F_a = \left(\frac{m_{ion}}{\tau_c}\right) 2a\nu\, exp\left(\frac{-W_a}{kT}\right) sinh\left(\frac{azeE_a}{2kT}\right) \quad (8)$$

where $a$ is the jump distance of the ions between hopping, $\nu$ is a frequency factor, $W_a$ is the energy barrier, $k$ is Boltzmann's constant, $T$ is absolute temperature, and $E_a$ is the electric field in the thin film. Combining (5)-(8) produces:

$$m_{ion}\frac{d^2x}{dt^2} + \frac{m_{ion}}{\tau_c}\frac{dx}{dt} + \frac{(ze)^2 N_d}{\epsilon_r \epsilon_0}(x - x_0) = \left(\frac{m_{ion}}{\tau_c}\right) 2a\nu\, exp\left(\frac{-W_a}{kT}\right) sinh\left(\frac{azeE_a}{2kT}\right) \quad (9)$$

For cases in which the thin film is bounded between electrodes separated by a distance $D$ and in which the ions are unable to penetrate the electrode a boundary condition $0 \leq x(t) \leq D$ may be added.

In order to analyze the collective motion of an aggregate number of ions in the thin films it will be necessary to modify (9). Fig. 2a illustrates a uniform distribution of ions within a region of a thin film having thickness D. If there is sufficient ionic conductivity in the medium the ions will gradually redistribute themselves due to electrostatic repulsion leading to Fig. 2b. However the total number of ions should not change due to conservation of charge[1].

$$N_d(t) x_{d0}(t) = N_d x_{d0} \quad (10)$$

As a voltage is applied to the thin film as in Fig. 2c the mean value of the ionic distribution drifts and is independent of redistribution of the collective ions. In light of the above discussion (9) may be rewritten in terms of the aggregate motion of the ions where $x_d(t)$ is the dynamic mean of the actual positions of the ions and $x_0(t)$ is the dynamic mean equilibrium[2] position of the ions.

$$m_{ion} \frac{d^2 x_d(t)}{dt^2} + \frac{m_{ion}}{\tau_c} \frac{dx_d(t)}{dt} + \frac{(ze)^2 N_d(t)}{\epsilon_r \epsilon_0}(x_d(t) - x_0(t)) = \left(\frac{m_{ion}}{\tau_c}\right) 2av \exp\left(\frac{-W_a}{kT}\right) \sinh\left(\frac{azeE_a}{2kT}\right) \quad (11)$$

This may be easily simplified to:

$$\frac{d^2 x_d(t)}{dt^2} + \frac{1}{\tau_c} \frac{dx_d(t)}{dt} + \frac{(ze)^2 N_d(t) x_d(t)}{m_{ion} \epsilon_r \epsilon_0}\left(1 - \frac{x_0(t)}{x_d(t)}\right) = \left(\frac{2av}{\tau_c}\right) \exp\left(\frac{-W_a}{kT}\right) \sinh\left(\frac{azeE_a(t)}{2kT}\right) \quad (12)$$

Further simplification of (12) may be achieved in the case when the aggregated ions are adjacent the left or right electrode. In Fig. 3a-c a dynamic evolution of the ion distribution is illustrated in a similar manner as in Fig. 2 in which the ions are accumulated at a boundary. In this case it is possible to relate the dynamic mean $x_d(t)$ with the dynamic width $x_{d0}(t)$. For a uniform ion distribution as illustrated $x_d(t) = x_{d0}(t)/2$. The dynamic equation (12) may now be further simplified using (10) as:

---

[1] It is notable that in the special case of solid electrolyte thin films ions are temporarily produced from an electrochemically active electrode and are critical to the formation of metal filaments important to resistance switching. This case will be discussed in section III in the sub-section on active junctions.

[2] The concept of "dynamic mean equilibrium" may seem confusing to some. It is "equilibrium" in the sense it is the mean position of the ions in steady state when the forcing function is zero. It is dynamic in the sense that a sufficiently high forcing function will change its value and transient dynamics will still be in play.

$$\frac{d^2 x_d(t)}{dt^2} + \frac{1}{\tau_c} \frac{dx_d(t)}{dt} + \frac{(ze)^2 N_d x_{d0}}{2 m_{ion} \epsilon_r \epsilon_0}\left(1 - \frac{x_0(t)}{x_d(t)}\right) = \left(\frac{2av}{\tau_c}\right) \exp\left(\frac{-W_a}{kT}\right) \sinh\left(\frac{azeE_a(t)}{2kT}\right) \quad (13)$$

Still further simplifications can be made under the assumptions that

a) the applied field $E_a(t)$ is sufficiently small that the mean equilibrium is time-independent ($x_0(t)=x_0=x_{d0}/2$) and the sinh function can be approximated as linear,

b) the variation $\Delta x_d(t)$ is small compared to $x_0$,

$$\Delta x_d(t) = x_d(t) - x_0 \ll x_0 \quad (14)$$

Under these conditions (13) may be approximated as:

$$\frac{d^2 \Delta x_d(t)}{dt^2} + \frac{1}{\tau_c} \frac{d\Delta x_d(t)}{dt} + \frac{(ze)^2 N_d}{m_{ion} \epsilon_r \epsilon_0} \Delta x_d(t) = \left(\frac{a^2 vze}{\tau_c kT}\right) \exp\left(\frac{-W_a}{kT}\right) E_a(t) \quad (15)$$

Finally we arrive at a tractable form! This is of course the familiar driven damped harmonic oscillator differential equation.

$$\frac{d^2 x}{dt^2} + 2\zeta\omega_0 \frac{dx}{dt} + \omega_0^2 x = F(t) \quad (16)$$

We can immediately obtain a potentially useful result in terms of the resonant angular frequency $\omega_0$.

$$\omega_0 = \sqrt{\frac{(ze)^2 N_d}{m_{ion} \epsilon_r \epsilon_0}} \quad (17)$$

By using (17) the effective mass of ions or oxygen vacancies in mem-resistor thin films might be able to be estimated from the frequency at which hysteresis effects are at a maximum.

III. COUPLING IONIC AND ELECTRONIC DYNAMICS

*a) Dynamic Schottky junction*

The behavior of electrons in semiconductor thin films is well studied and reviewed such as in [14].

In the case of metal-semiconductor junctions the situation is such that if the work function of the semiconductor is smaller than the metal the electrons in the semiconductor are more energetic and become depleted from the semiconductor side of the barrier while accumulating on the metal side. This situation is referred to as a Schottky barrier and is illustrated by the left side of Fig. 4a-c.

In the case where the work function of the metal is less than the semiconductor the electrons in the semiconductor are less energetic and electrons enter into the semiconductor from the metal. This situation is referred to as an ohmic contact and is illustrated by the right side of Fig. 4a-c.

The concept of built-in voltage $\phi_i$ is used in reference to metal-semiconductor junctions referring to the difference between the work functions of metal $\Phi_M$ and that of a semiconductor $\Phi_S$. In general both the built-in voltage and semiconductor work functions can be dependent on position *and time* as determined by the dynamic evolution of ionic density.

$$\phi_i(x,t) = \Phi_M - \Phi_S(x,t) \quad (18)$$

If the dynamic ionic density $N_d(x,t)$ is known Poisson's equation can be used to determine this built-in voltage.

$$\frac{d^2\phi(x,t)}{dx^2} = -\frac{ze}{\epsilon_r\epsilon_0} N_d(x,t) \quad (19)$$

Assume (as illustrated in Fig. 4b) that at an initial time the ionic doping $N_d$ present near the left junction (0<x<$x_{d0}(t)$) is constant. Poisson's equation in this region reduces to

$$\frac{d^2\phi(x,t)}{dx^2} = -\frac{ze}{\epsilon_r\epsilon_0} N_d(t) \quad (20)$$

This equation can be used to solve for the electric field $E(x,t)$ and potential $\phi(x,t)$ given the conditions $E(x_{do}(t),t)=0$ and $\phi(0,t)=0$.

$$E(x,t) = \frac{d\phi_i(t)}{dx} = \frac{-ze}{\epsilon_r\epsilon_0} N_d(t)(x_{d0}(t)-x) \quad (21)$$

$$\phi(x,t) = \frac{-ze}{\epsilon_r\epsilon_0} N_{d0}(t)(x_{d0}(t)x - x^2/2) \quad (22)$$

The built-in voltage $\phi_i$ is defined at the boundary of the electron depletion region x=$x_{d0}$.

$$\phi_i(t) = |\phi(x_{d0},t)| = \frac{ze}{2\epsilon_r\epsilon_0} N_d(t) x_{d0}^2(t) \quad (23)$$

Given (10) this can be simplified to

$$\phi_i(t) = \frac{ze}{2\epsilon_r\epsilon_0}(N_{d0}x_{d0})x_{d0}(t) \quad (24)$$

The harmonic equation (15) can now be used to determine the time-dependent behavior of the built in voltage with the dynamic mean equilibrium located at $x_0=x_{d0}/2$ and noting that the variation of the depletion width is twice the variation of the dynamic mean.

$$x_{d0}(t) = x_{d0} + 2\Delta x_d(t) \quad (25)$$

This can be applied to a Schottky barrier in which the relationship between the electron current density $J(t)$ and applied voltage $V_a(t)$ is expressed as

$$J(t) = J_S(t)\left[\exp\left(\frac{zeV_a(t)}{kT}\right) - 1\right] \quad (26)$$

$$J_S(t) = \frac{(ze)^2 D_n N_c}{kT}\left[\frac{2ze(\phi_i(t)-V_a(t))N_d(t)}{\epsilon_r\epsilon_0}\right]^{1/2} \exp\left(\frac{-ze\phi_B}{kT}\right) \quad (27)$$

Using (10), (24), and (25) the Schottky barrier equations (26), (27) may be expressed in terms of the single dynamic variable $\Delta x_d(t)$ and the applied voltage $V_a(t)$ with the other quantities being time-independent.

In addition to the current-voltage relationship of the Schottky barrier, capacitive current density $J_c(t)$ can be calculated using the dynamic capacitance $C(t)$ expressed in terms of the depletion width $x_{d0}(t)$.

$$J_c(t) = \frac{d[C(t)V_a(t)]}{dt} = \frac{d}{dt}\left[\frac{\epsilon_r\epsilon_0}{x_{d0}(t)} V_a(t)\right] \quad (28)$$

The equations for a dynamic Schottky barrier are summarized in TABLE 1.

*b) Dynamic tunneling junction*

Memory resistors have been under development by a company called Unity Semiconductor since 2002 based on a conductive metal oxide having a variable tunneling barrier [15]. Fig. 5a provides an approximate energy diagram for a tunneling barrier formed by an ion-depleted non-conductive region of thickness $x_{d0}(t)$ wherein $x_{d0}(t)$ is sufficiently small to allow electron tunneling. The ion depleted region separates the left electrode from a region having a uniform ion density $N_{d0}(t)$ which may be approximated as time independent for $x_{d0}(t) \ll D$ under the condition of (10).

When the applied voltage to this system is zero ($V_a(t)=0$) the magnitude of the energy barrier $\Phi_B$ is the product of the constant electric field $E_0$ and the ion depletion width $x_{d0}(t)$.

$$\Phi_{B0}(t) = E_0 x_{d0}(t) \qquad (29)$$

At zero voltage some tunneling between the metal and ionic region occurs due to the thermal energy of the electrons. At equilibrium the tunneling current density $J_{T0}$ from the metal to the ionic region should balance the tunneling current density from the ionic region to the metal and can be calculated using the tunneling current equation referenced in [15]. Note that $x_{d0}(t)$ is time-dependent but is a constant for purposes of the integration with respect to x.

$$J_{T0}(t) = C_0 \exp\left(\frac{-\sqrt{8m_e}}{h/2\pi} \int_0^{x_{d0}(t)} \sqrt{\Phi_{B0}(t)} dx\right) = C_0 \exp\left(\frac{-\sqrt{8m_e}}{h/2\pi} \sqrt{E_0(t) x_{d0}^3(t)}\right) \qquad (30)$$

As a positive voltage is applied to the left electrode the height of the barrier decreases so that

$$\Phi_{Bv}(t) = E_0 x_{d0}(t) - V_a(t) \qquad (31)$$

and the tunneling current is now calculated as

$$J_{Tv}(t) = C_0 \exp\left(\frac{-\sqrt{8m_e}}{h/2\pi} \int_0^{x_{d0}(t)} \sqrt{\Phi_{Bv}(t)} dx\right) = C_0 \exp\left(\frac{-\sqrt{8m_e}}{h/2\pi} \sqrt{E_0(t) x_{d0}^3(t) - V_a(t) x_{d0}^2(t)}\right) \qquad (32)$$

The net increase of current from equilibrium is

$$J_T(t) = J_{Tv}(x_{d0}(t)) - J_{T0}(x_{d0}(t)) \qquad (33)$$

For $x_{d0}(t) \ll D$ the dynamic mean equilibrium $x_0$ of the ionic region is approximately D/2 and the variation of the ionic depletion width $x_{d0}(t)$ is twice the variation of the dynamic mean.

$$x_{d0}(t) = x_{d0} + 2\Delta x_d(t) \qquad (34)$$

The harmonic equation (15) can now be used to determine the time-dependent behavior of the tunneling barrier (with $x_0=D/2$).

$$\frac{d^2 \Delta x_d(t)}{dt^2} + \frac{1}{\tau_c} \frac{d\Delta x_d(t)}{dt} + \frac{(ze)^2 N_d}{m_{ion} \epsilon_r \epsilon_0} \Delta x_d(t) = -\left(\frac{a^2 vze}{\tau_c kT}\right) \exp\left(\frac{-W_a}{kT}\right) V_a(t)/x_{d0} \qquad (35)$$

It is expected in this example the dynamic capacitance would be negligible since the right junction is ohmic and the left junction is tunneling. The equations for such a dynamic tunneling barrier are summarized in TABLE 1.

*c) Generalized dynamic diode and tunnel junctions*

It is to be expected that in the case of a non-uniform ion distributions $N_d(x,t)$ or in consideration of more detailed quantum mechanical models the dynamic Schottky and tunneling mem-resistor equations derived above will need to be modified. In general one may discuss the mem-resistor model in terms of an electron current density expressed as the difference between the diffusion or tunneling current when $V_a(t)=0$ and the current when $V_a(t) \neq 0$.

$$J(t) = J_0\big(\phi_i(t), N_d(t), x_{d0}(t), V_a(t)\big) - J_0\big(\phi_i(t), N_d(t), x_{d0}(t), 0\big) \qquad (36)$$

A generalized version of (10) may be formulated for non-uniform ionic density.

$$\int_0^{x_{d0}(t)} N_d(x,t) dx = N_{da} x_{d0} \qquad (37)$$

In this case $N_{da}$ represents the spatial average of the ionic density in the depletion region and $x_{d0}(t)$ is the dynamic deletion width. The product of these values is time–independent provided that the number of ions is conserved. When the equilibrium ionic density is known integrating (38) provides a relation between $x_{d0}(t)$ and $N_d(x,t)$ and a similar analysis as in Section II can be performed.

Poisson's equation can also be solved given the non-uniform density $N_d(x,t)$ as in section IIIa and this can be used to form a relationship between $N_d(x,t)$ and the dynamic built-in voltage $\phi_i(t)$. These relationships can be used to express (37) as a function of a single dynamic variable.

$$J(t)=J_0(x_{d0}(t),V_a(t)) - J_0(x_{d0}(t),0) \quad (38)$$

This equation combined with the appropriately modified version of (12) result in a general dynamic description of diodes and tunnel junctions.

*d) Frequency Response*

In the case of a sinusoidal voltage applied to the dynamic tunneling junction

$$E_a(t) = \frac{V_0}{x_{d0}} sin(\omega t) \quad (39)$$

the steady-state solution to (36) takes the form

$$\Delta x_{d0}(t) = \Delta X_d sin(\omega t + \varphi_0) \quad (40)$$

$$\Delta X_d = \frac{a^2 \nu z e \; exp\left(\frac{-W_a}{kT}\right) V_0 / x_{d0}}{\tau_c kT \sqrt{(\omega/\tau_c)^2 + (\omega^2 - \frac{(ze)^2 N_d}{m_{ion} \epsilon_r \epsilon_0})^2}} \quad (41)$$

$$\varphi_0 = \tan^{-1} \frac{\omega}{(\omega^2 - \frac{(ze)^2 N_d}{m_{ion} \epsilon_r \epsilon_0})\tau_c} \quad (42)$$

It is notable that at resonance $\varphi_0 = 90$ degrees and the dynamic behavior of the tunneling width is 90 degrees out of phase with the applied voltage. As a result a zero-crossing hysteresis curve will develop in the current vs. voltage curve. As the input signal frequency increases or decreases sufficiently from the resonance frequency the phase shift $\varphi_0$ will go to zero and the hysteresis effect will disappear. It is these effects which led the HPLabs researchers to incorrectly come to the conclusion that they found a memristor. However, zero-crossing hysteresis is a necessary but not a sufficient condition for Chua's memristor.

*e) Active junctions*

In some cases (10) (or the more generalized (37)) may not hold. For example in solid electrolyte memory electrochemical reactions occur at active electrodes which temporarily create ions during electroformation of a filament [13]. In another example electrons may be temporarily trapped at an interface at a related rate to an applied periodic voltage as in [11]. In these cases (10) would need to be modified according to

$$N_d(t) x_{d0}(t) = N_d x_{d0} + Q_S(t)/ze \quad (43)$$

where $Q_S(t)$ is the dynamic charge density and is related to the electrochemical reaction rate or probability of charge trapping. In cases where the dynamic charge density is significant non-zero crossing hysteresis effects will emerge in the I-V curve and effects such as illustrated in Fig. 1a of Argall [4] or in [11] will emerge. Following through with the previous analysis of Section II (13) may be adjusted to incorporate active ionic mem-resistors.

$$\frac{d^2 x_d(t)}{dt^2} + \frac{1}{\tau_c} \frac{dx_d(t)}{dt} + \frac{(ze)^2 (N_d x_{d0} + Q_S(t)/ze)}{2 m_{ion} \epsilon_r \epsilon_0}\left(1 - \frac{x_0(t)}{x_d(t)}\right)$$
$$= \left(\frac{2a\nu}{\tau_c}\right) exp\left(\frac{-W_a}{kT}\right) sinh\left(\frac{aze E_a(t)}{2kT}\right) \quad (44)$$

IV. CONCLUSION

Researchers at HPLabs have used the concept of Chua's memristor [1] in a propaganda campaign to make it appear as if they discovered something new. However, the magnetic flux vs. charge relationship of HP's memristor [2] is actually useless to the analysis of real nanoionic thin films since it neglects effects such as the inertia of ions, repulsive force between ions, and capacitance changes due to ion motion in thin films. Since the 1990's numerous competitors to HP such as Advanced Micro Devices, Axon Technologies, Micron Technologies, Samsung, Sharp, and Unity Semiconductor have been developing ReRAM based on thin film ionic materials [5]. Meanwhile, the researchers at HPLabs were trying to develop a form of molecular memory. I believe that when the researchers at HPLabs found that there was more progress being made by their competitors they decided to switch their research efforts away from molecular materials and toward ionic thin films. In order to give the appearance that they were an initiator rather than copying their competitors the researchers of HPLabs decided to use the idea of Chua's memristor as a "fourth fundamental circuit element" which was allegedly missing until HP's "discovery".

The use of TiO$_2$ as a resistance switching material was known since the 1960's [4] and the concept of a memory resistor predates Chua [3]. There is no realistic model of ionic thin films based on the magnetic flux vs. charge relationship of a memristor. Chua's memristor remains nothing more than a mythological construct [6]. It is not practical to form the basis of new technology on mythology.

In contrast this paper has provided a realistic analysis of resistance switching of nanoionic thin films grounded in the physics of ions and semiconductors. It is hoped that the equations summarized in TABLE 1 will be of assistance to further development of ReRAM. It is also hoped that they will assist to further develop my patented inventions involving mem-resistor crossbars used in signal processing circuits and robotic control systems [16].

It is requested that honest researchers who work on ReRAM or dynamic circuit modeling should refer to any devices which conform to the equations described herein as *mem-resistors* to distinguish these devices from the propaganda of the magnetic flux vs. charge memristor of Chua.

*TABLE 1 Summary of Mem-resistor Equations*

## A) Dynamic Schottky junction

$$J(t) = J_S(t)\left[\exp\left(\frac{zeV_a(t)}{kT}\right) - 1\right]$$

$$J_S(t) = \frac{(ze)^2 D_n N_c}{kT}\left[\frac{2ze\left(\frac{ze}{2\epsilon_r\epsilon_0}\right)(N_{d0}x_{d0}(x_{d0}+2\Delta x_d(t)) - V_a(t))N_{d0}x_{d0}}{\epsilon_r\epsilon_0(x_{d0}+2\Delta x_d(t))}\right]^{1/2} \exp\left(\frac{-ze\phi_B}{kT}\right)$$

$$J_c(t) = \frac{d[C(t)V_a(t)]}{dt} = \frac{d}{dt}\left[\frac{\epsilon_r\epsilon_0}{x_{d0}+2\Delta x_d(t)}V_a(t)\right]$$

$$\frac{d^2\Delta x_d(t)}{dt^2} + \frac{1}{\tau_c}\frac{d\Delta x_d(t)}{dt} + \frac{(ze)^2 N_d}{m_{ion}\epsilon_r\epsilon_0}\Delta x_d(t) = -\left(\frac{a^2\nu ze}{\tau_c kT}\right)\exp\left(\frac{W_a}{kT}\right)\frac{V_a(t)}{x_{d0}}$$

## B) Dynamic Tunneling junction

$$J_T(t) = C_0\left(\exp\left(\frac{-\sqrt{8m_e}}{h/2\pi}\sqrt{E_0(t)x_{d0}^3(t) - V_a(t)x_{d0}^2(t)}\right) - \exp\left(\frac{-\sqrt{8m_e}}{h/2\pi}\sqrt{E_0(t)x_{d0}^3(t)}\right)\right)$$

$$\frac{d^2\Delta x_d(t)}{dt^2} + \frac{1}{\tau_c}\frac{d\Delta x_d(t)}{dt} + \frac{(ze)^2 N_d}{m_{ion}\epsilon_r\epsilon_0}\Delta x_d(t) = -\left(\frac{a^2\nu ze}{\tau_c kT}\right)\exp\left(\frac{W_a}{kT}\right)\frac{V_a(t)}{x_{d0}}$$

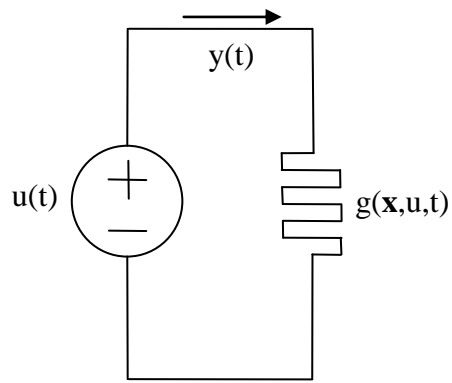
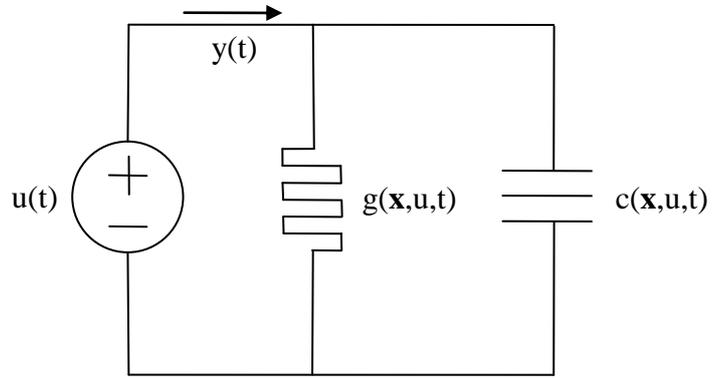

a) Memristive System

b) Mem-Admittance System

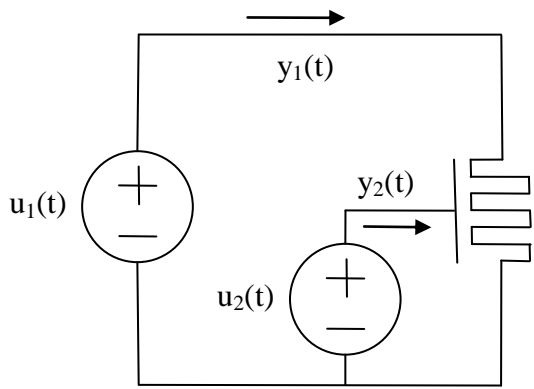
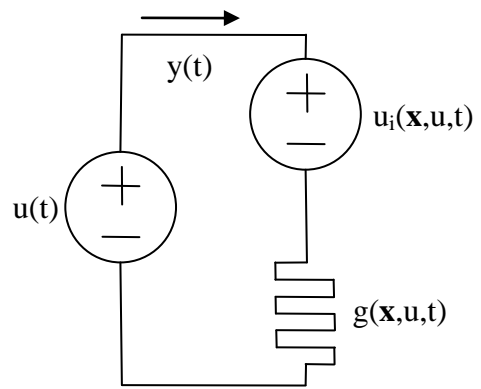

c) Mem-Transistor System

d) Mem-Active System

Fig. 1 Different types of Mem-Electronic Systems

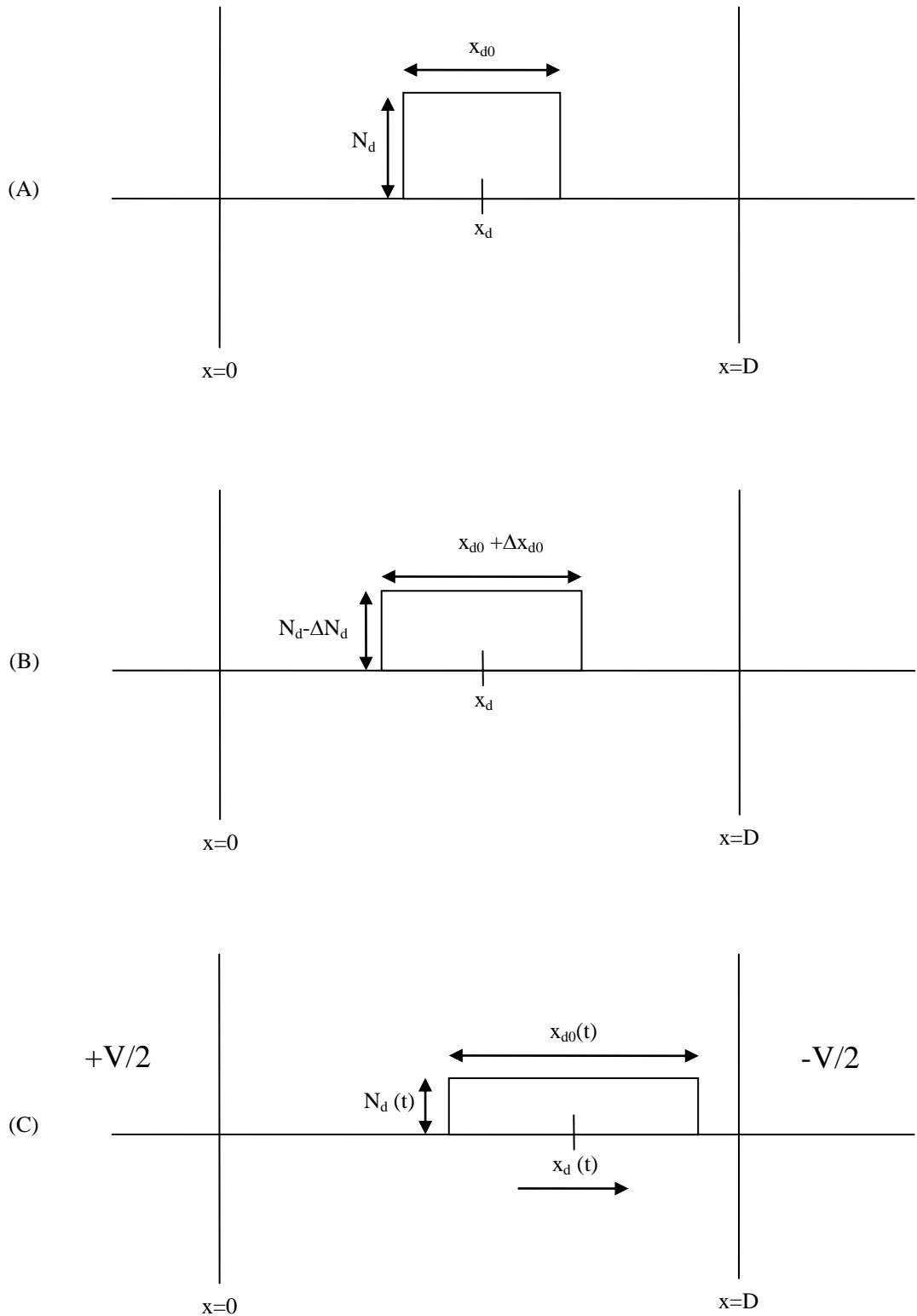

Fig. 2
(A) Uniform positive ion distribution at initial time with zero applied voltage. The mean value of the distribution is located at $x_d$.
(B) At later time ions spread out but charge is conserved. $(N_d-\Delta N_d)(x_{d0}+\Delta x_{d0})=N_d x_{d0}$.
(C) A voltage is applied moving the mean of the ions toward the right. While the distribution $N_d(t)$ may change charge is conserved.

June 25, 2011 (ver.3)

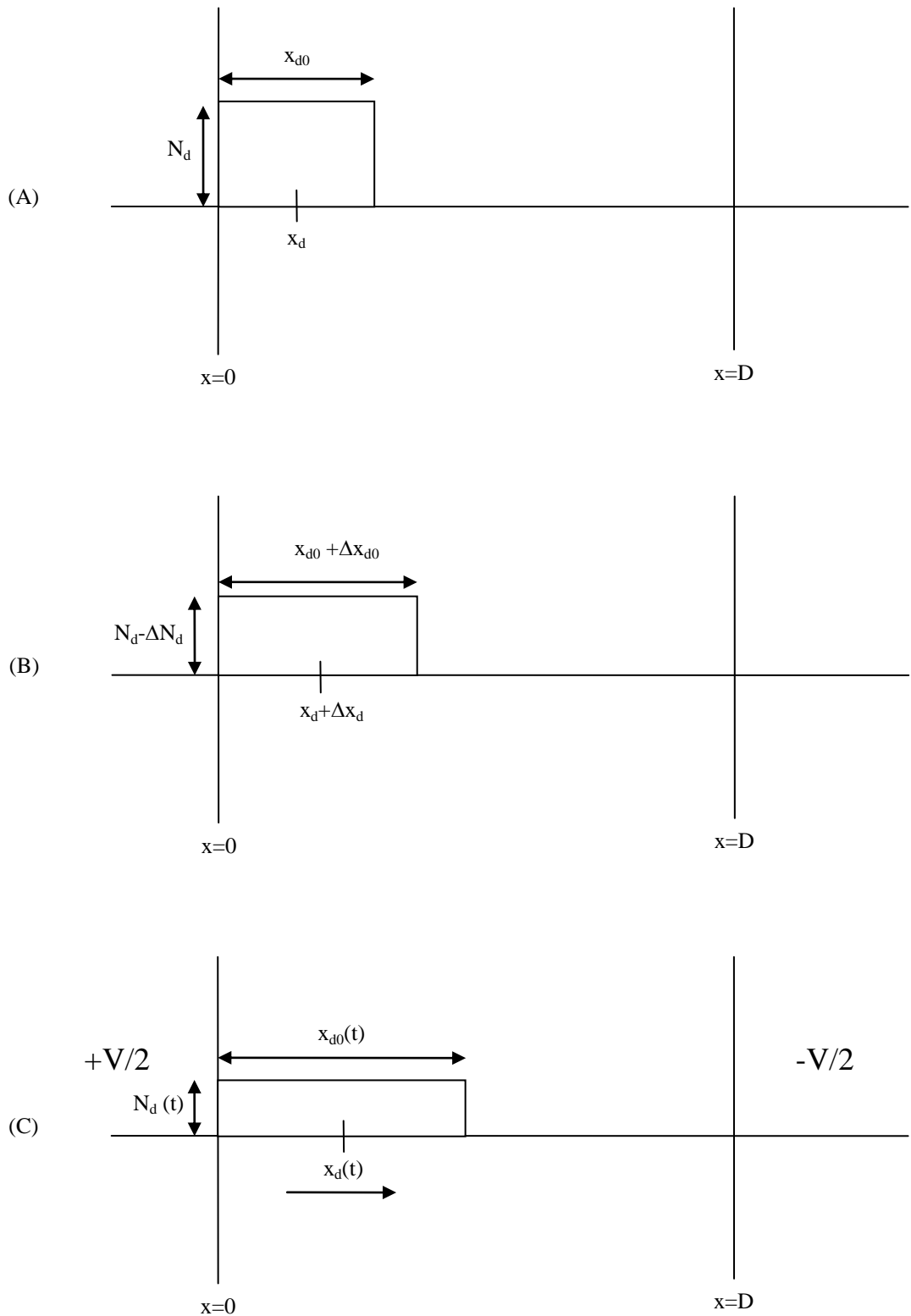

Fig. 3
(A) Uniform positive ion distribution at initial time with zero applied voltage. The mean of the distribution is located at $x_d$.
(B) At later time ions spread out but charge is conserved. $(N_d - \Delta N_d)(x_{d0} + \Delta x_{d0}) = N_d x_{d0}$.
(C) A voltage is applied moving the mean of the ions toward the right. While the distribution $N_d(t)$ may change charge is conserved.

June 25, 2011 (ver.3)

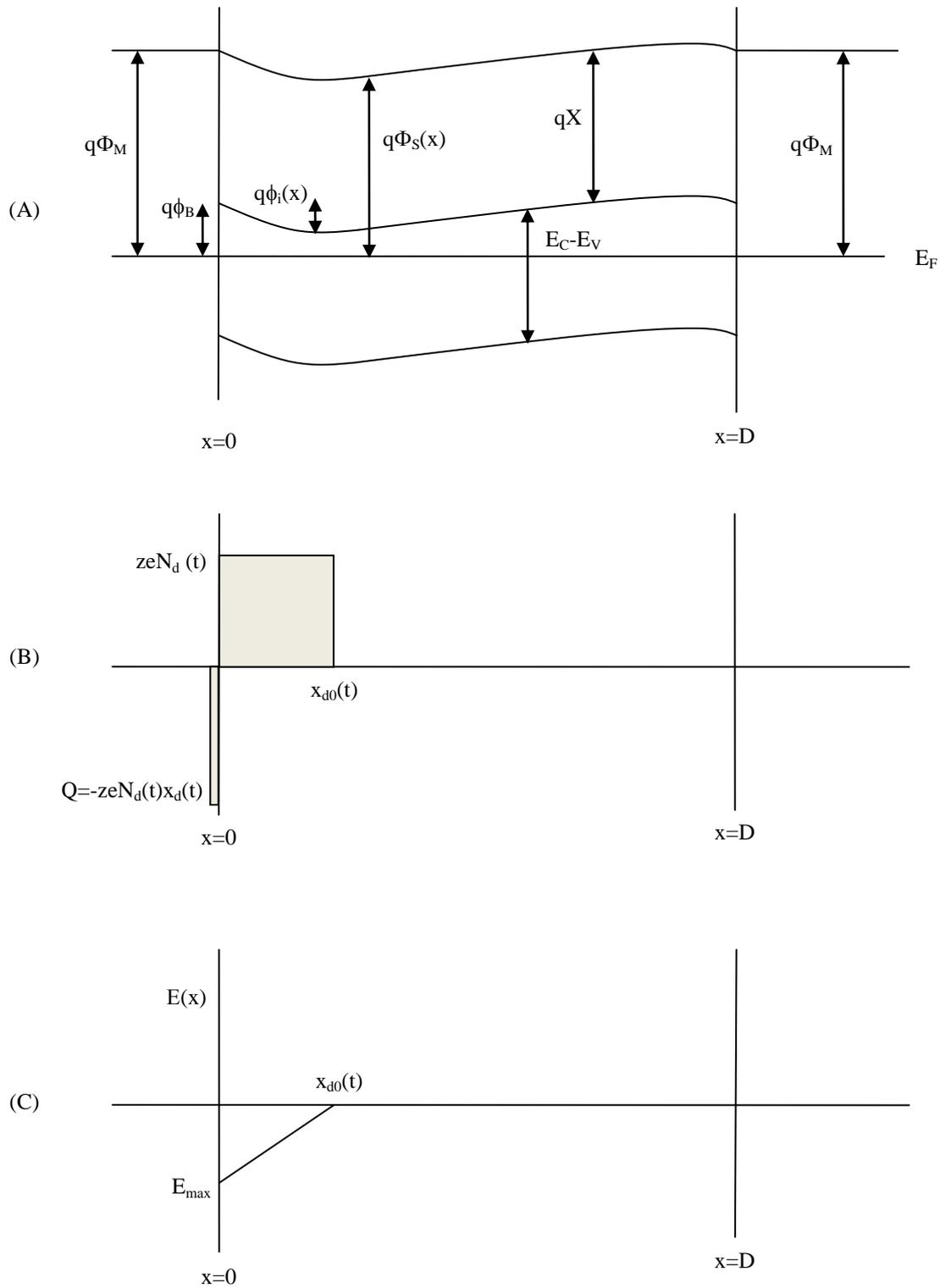

Fig. 4
(A) Idealized band diagram for a Schottky mem-resistor under depletion approximation.
(B) Charge density for the Schottky mem-resistor which illustrates electron depletion region located at $0 \leq x \leq x_{d0}$. Outside depletion region $N_d \approx n$ and the ion charge is neutralized by the electron charge.
(C) Electric field in the device.

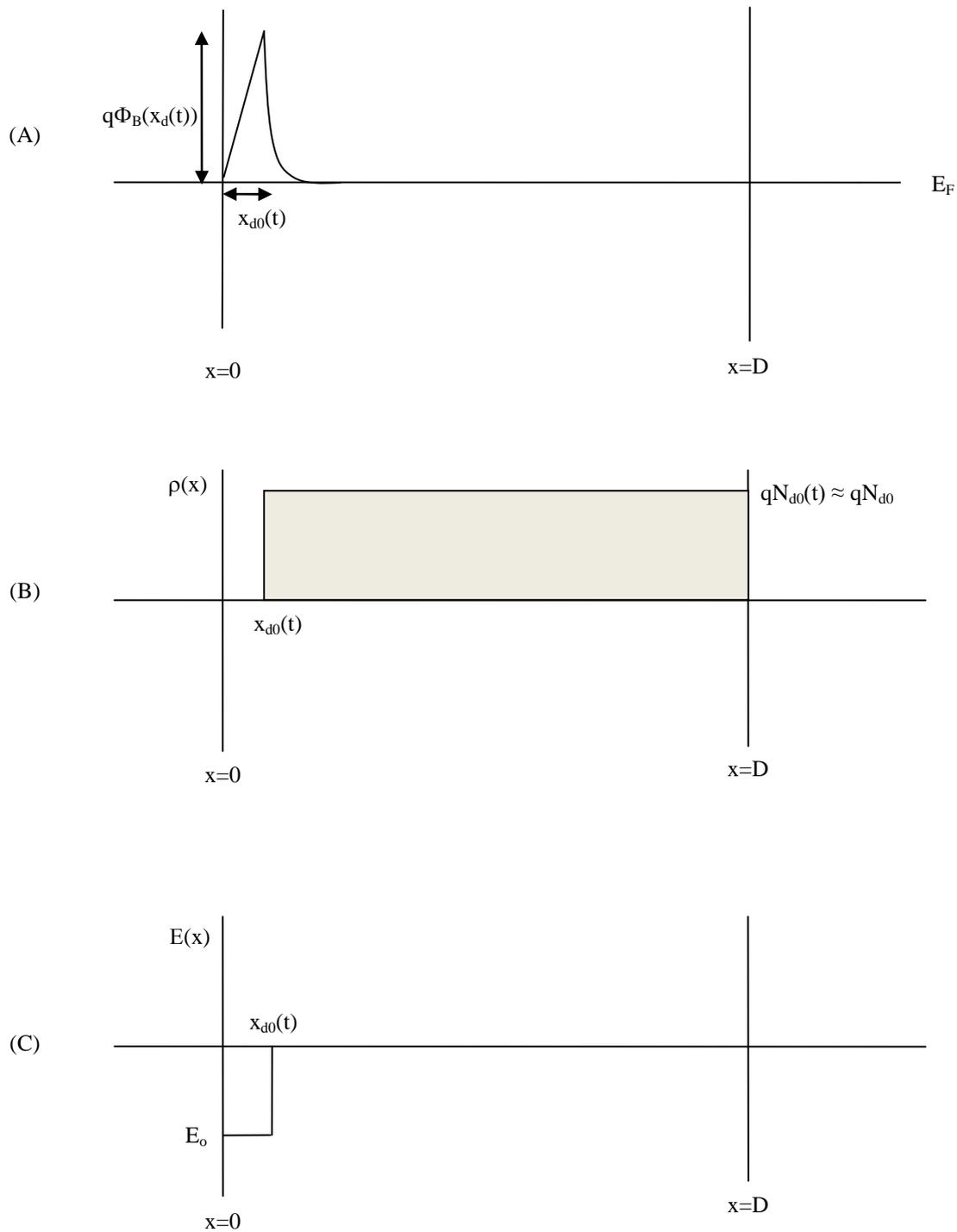

Fig. 5
(A) Idealized energy diagram for a tunneling mem-resistor.
(B) Charge density for tunneling mem-resistor having dynamic ionic depletion region $0<x<x_d(t)$.
(C) Electric field in the tunneling mem-resistor.